# Self-Biased Integrated Magnonic Device

M. Cocconcelli*[1], F. Maspero[1], A. Micelli[1], A. Toniato[1], A. Del Giacco[1], N. Pellizzi[1], A. E. Plaza[1], A. Cattoni[1], M. Madami[2], R. Silvani[2], A. A. Hamadeh[3], C. Adelmann[4], P. Pirro[3], S. Tacchi[5], F. Ciubotaru[4], R. Bertacco*[1]

[1] Dipartimento di Fisica, Politecnico di Milano, Via G. Colombo 81, 20133 Milano (Italy)
[2] Dipartimento di Fisica e Geologia, Università di Perugia, Via A. Pascoli, 06123 Perugia, Italy
[3] Fachbereich Physik and Landesforschungszentrum OPTIMAS, Rheinland-Pfälzische Technische Universität Kaiserslautern-Landau, 67663 Kaiserslautern, Germany
[4] imec, Kapeldreef 75, 3001 Heverlee, Belgium
[5] Istituto Officina dei Materiali del CNR (CNR-IOM), Unità di Perugia, c/o Dipartimento di Fisica e Geologia, Università di Perugia, Via A. Pascoli, 06123 Perugia, Italy

* maria.cocconcelli@polimi.it, riccardo.bertacco@polimi.it

In the race towards "beyond 6G" telecommunication platforms, magnonics emerges as a promising solution due to its wide tunability within the FR3 band (7–24 GHz). So far, however, the need of an external magnetic bias field to allow the coherent excitation of spin waves has been a major bottleneck. Conventional bulky electromagnets are power-intensive and challenging to integrate on-chip, restricting magnonic applications largely to academic research.

Here, we present the first demonstration of a standalone, tunable magnonic device featuring all-electric input and output, fully integrated on a silicon substrate with a compact footprint of $100 \times 150$ μm². The device consists of a CoFeB waveguide equipped with two radio frequency antennas, flanked by a symmetric configuration of T-shaped magnetic flux concentrators and rectangular SmCo permanent micromagnets. By varying the distance D between the flux concentrators and the permanent magnets from 0 to 12 μm, the transverse bias field can be tuned from 20.5 mT to 11 mT, respectively. This variation directly modulates the dispersion relation of Damon-Eshbach spin wave modes in the CoFeB waveguide.

In these proof-of-concept devices, the spin wave frequency band ranges from 3 to 8 GHz, with precise phase shift tuning of up to 120° at 6 GHz achieved by varying D within the 0–8 μm range. The operational frequency band could even be pushed to higher frequencies through optimized micromagnet engineering.

# 1. INTRODUCTION

So far magnonics, i.e. the branch of magnetism studying spin waves (SW) in magnetic media, essentially remained an exciting academic research field dealing with this peculiar form of electromagnetic waves in ferro/ferri/antiferro-magnetic materials displaying intriguing phenomena [1][2][3][4]. At variance with other forms of waves, the band-dispersion can be easily tuned by setting the relative orientation between the magnetic bias field (H0) and the wave vector (k) which defines the different configurations: "Forward Volume - FW" (H0 out-of-plane, k in-plane), "Backward volume - BV" (H0 //k , in-plane), "Damon Eshbach – DE" (H0 $\perp$k, in-plane) [5]. Furthermore, within each configuration, the band dispersion can be finely tuned by varying the magnitude of H0, thus providing a "easy" reconfigurability of a magnonic device. Beyond the linear regime, SW display intriguing non-linear processes connected to multi-magnon scattering etc[6]. Bose condensation (BC) in BV configurations has been reported and used to investigate intriguing properties of BC states[7]. Furthermore, the interaction between SW and magnetic media can be exploited to move domain walls and reconfigure memory/logic elements. Finally, SW "naturally" provide a viable alternative to acoustic waves for "Beyond 6G" applications. While surface and bulk acoustic wave technology is facing challenges in covering the demands of a reliable operation over the whole frequency ranges involved while maintaining reasonable insertion losses, reduced complexity and low power dissipation, spin waves can cover a broad spectrum, ideally from a few MHz up to 1 THz. This makes them a promising candidate for the implementation of high-frequency communication technology [8][9][10]. However, just a few examples of applications of magnonics exist on the market. Apart from Yttrium-Iron-Garnet (YIG) tuned RF oscillators and Auto-tune filters used to mitigate electro-magnetic interference (e.g. by Metamagnetics Inc.) there are no real implementations of exciting devices reported in literature such as RF filters, phase-shifters, interferometers, spectrum analyzers, etc. The main reason for this is twofold: (i) many proposed devices are based on epitaxial YIG, providing the longest propagation length but not suitable for integration with Silicon electronics, (ii) to achieve high frequencies (above 4-5 GHz) interesting for typical applications, magnonic devices require an external magnetic bias field. Bulky electromagnets used in laboratory to produce such a field are not compatible with consumer electronic devices.

In this paper we address these two bottlenecks by providing a first demonstration of a standalone magnonic device fully integrated on silicon, not requiring any external magnetic field bias for its operation and suitable for application in portable devices like mobile phones. Even though there are some papers reporting zero-field propagation of spin waves in nanostructures where single domain configuration can be set by shape anisotropy, and other examples of system-integration of magnonic-based RF devices and electromagnets (or permanent magnets), to the best of our knowledge this work represents the first example of monolithic integration on Silicon [11][12].

Our concept is based on the exploitation of a suitable combination of permanent micro-magnets and magnetic flux concentrators (MFCs), fabricated on the same silicon chip hosting the magnonic waveguide, to produce a bias field up to 20.5 mT which is strong enough to stabilize a DE configuration. Varying the distance (D) between the permanent micromagnets and the MFC we can tune the bias field in the 11-20.5 mT range so that our first proof-of-concept device behaves like a tunable phase-shifter in the 3-8 GHz range, even though the frequency band can be easily shifted upwards by proper engineering of the permanent magnets and MFCs. The device is equipped with two (input – output) RF antennas for a full electric operation of the device whose total footprint (100x150 μm) makes it compatible with a system on chip or monolithic integration with CMOS electronics.

## 2. RESULTS

### 2.1. Chip layout

Figure 1a shows a false-color SEM image of our device. The green vertical rectangle represents the CoFeB waveguide (25 nm thick, 3.6 μm wide) for SW propagation, with two inductive antennas (grey) made of gold (125 nm thickness, 1 μm width) above the CoFeB for the input and output RF signals. Between the antennas and the CoFeB conduit there is a silica spacer with thickness $t_{spacer}$ of about 70 nm. The central region of the conduit, with the two antennas at 5 μm distance (center to center), is flanked by two T-shaped MFC (blue), made of a Py/Cr multilayer with tapered edges, having 1 μm thickness and 10 μm width (see methods for details). Two SmCo permanent micromagnets (pink rectangles in Figure 1) with remanent magnetization $\mu_0 M_R$ = 0.6 T upon magnetization at 2 T, 1 μm thickness and 50x100 μm footprint, are placed at a variable distance D (0, 1.5, 4, 8, 12 μm) from the MFC in order to tune the coupling between them and, consequently, the field $H_0$ produced in the CoFeB region between the antennas [13][14][15][16]. The MFC action is visible in Figure 1b, showing the results of a COMSOL simulation of the static magnetic field for a distance D = 3 μm. More details on the field profiles can be found in figure S1 of the Supporting Information. From numerical simulations we know that the MFC allows to concentrate the field to gain a factor G > 2 with respect to an external uniform bias field, within the linear region of the MFC (about +- 5 mT). On the other hand, G can become lower than unity for larger applied fields, which induce a partial saturation of the MFC. Note that the T-shaped configuration for the MFC has been chosen to maximize the bias magnetic field produced between their poles under the action of the stray field from the SmCo permanent magnets, which is not uniform at all. For this reason, the T-shape has been chosen instead of more common trapezoidal shapes which can provide higher gain for uniform fields but revealed to be less efficient when coupled to rectangular SmCo magnets (see Section 1 of the Supporting Information).

We carried out some microMOKE experiments to first characterize the MFC behavior and estimate the bias field $H_0$ produced at the center of the CoFeB waveguide for the various distances D. Figure 1c shows the loops measured for D = 0 in the pristine state of the SmCo permanent magnet, i.e. prior to their magnetization with a field of 2 T, with an applied transverse field Ha along the x direction. Blue and red curves represent, respectively, the loops taken at center of the CoFeB conduit, in the region flanked by the MFC, and at the top edge, far away from the horizonal arms of the MFC. We observe the typical "loop without hysteresis" expected for a transverse applied field with respect to the easy y-axis determined by shape anisotropy. The curves are quite symmetric with respect to Ha = 0, thus indicating that SmCo magnets in the pristine state do not provide a sizable bias. On the other hand, when moving from the edge to the center, the slope of the linear region increases by a factor 2.4±0.3 which represents a good estimate of the MFC gain (G) if we assume that the action of the MFC at the conduit edge is almost negligible, as seen in simulations. The situation completely changes upon magnetization of the SmCo magnet along the x-axis with Ha = 2 T, the maximum field available in our laboratory, as reported in panel 1d showing loops taken at the conduit center (loops taken at the edge are shown in Section 2 of the Supporting Information). A true hysteresis loop appears (blue line, corresponding to D = 0 μm), shifted by 9 mT towards negative fields because of the bias field from SmCo magnets. The reason for the appearance of hysteresis is connected to domain wall pinning by the non-uniform bias field in the central region of the conduit, as visible in the sequence of images form a video (Movie 1, available online as supporting material) reported in panels f,g,h,i of Figure 1, corresponding to snapshots taken at +7, -9.5, -10, -21 mT, respectively. We observe that magnetization reversal takes place via nucleation and propagation of domain walls which are pinned by the high field region between the horizontal arms of the MFC (see panel 1h) thus leading to the characteristic Barkhausen jumps seen in the loops and to hysteresis. Noteworthy, the shift of the loop ΔHa decreases from 9±0.2 to 6.2±0.2 mT when increasing D due to the less efficient coupling between the stray field from SmCo magnets and MFCs, as summarized in panel 1e, and this provides the way to tune the bias field $H_0$. Notice that $H_0$ does not coincide with the shift of the loop measured by MicroMOKE. Considering the gain of the MFC, $H_0$ = G·ΔHa, so that MicroMOKE provides a first estimate of the effective bias field achievable in our device: from 21.6±3.4 to 15±2.4 mT when varying D from 0 to 8 μm.

*2.2 Broadband spectroscopy*

The investigation of SW propagation in the DE configuration has been carried out using VNA spectroscopy.[17] First, we analyzed SW transmission in reference devices made of just the CoFeB waveguide and inductive antennas, by measuring the scattering matrix $S_{ij}$ as a function of frequency for different applied transverse fields $Ha_x$, from -100 mT to +100 mT in steps of 1 mT. The color plot of the imaginary part of $S_{12}$ vs. frequency and field is reported in Figure 2a, after subtraction of a reference signal taken at zero applied field Ha, when no coherent propagation of SW is expected, and noise removal via time gating (details on the signal processing are reported in the methods and in Section 3 of Supporting information). At low field the micromagnetic configuration of the conduit is not well defined so that we do not observe oscillations in the $Im(S_{12})$ which are the fingerprint of SW propagation. For |Ha| > 10 mT we start observing some oscillations, but a clear signature of DE modes appears only for |Ha| > 20 mT, where we see that the onset frequency of oscillations (close to the ferromagnetic resonance (FMR) frequency for the conduit at that peculiar field) follows a Kittel-like trend for a transversally applied external field. This is in good agreement with the micromagnetic simulations of the conduit without surrounding magnets (see Figure 3), showing that a single domain configuration with transverse magnetization is retrieved only above 20 mT of applied field.

The color plot of Figure 2a is essentially symmetric, apart from some small deviations at low fields which can be ascribed to small misalignments of the conduit with respect to the field [18] and from a general higher intensity of the oscillations for positive applied fields due to the non-reciprocal behavior of DE spin waves. It should be noted that there are two minima in the color maps of Figure 2a, at -12mT and +18mT, delimiting a region around Ha = 0 where the iso-phase lines display a slope with different sign with respect to the outer region. This is the typical hard-axis behavior indicating a reorientation of the sample magnetization. On reducing the strength of the external field, the magnetization does not remain oriented along the short axis of the waveguide, corresponding to a hard direction due to the shape anisotropy, and gradually rotates towards the long axis, leading to an increase of the spin wave frequency. Restricting our attention to positive fields, the outer region with positive slope can be associated to well-defined DE modes with positive group velocity, while the inner one reflects other BV-like modes with negative group velocity.

To gain a deeper insight into SW propagation in the conduits, in Figure 2b we report the traces of $Im(S_{12})$ for some selected applied fields (continuous lines), together with a fit (dashed lines) carried out with the following function:

$Im(S_{12}(\omega)) = S \cdot sin(k(\omega)r+\varphi) \cdot exp(-r/L_{att}) \cdot \eta(k(\omega))$ (1)

where S is the amplitude of the oscillating term $sin(k(\omega)r+\varphi)$ describing SW propagation, $k(\omega)$ is the inverse function of the band dispersion for DE modes in a metallic stripe, [19] [20] [21] *r* is the distance between input and output antennas, φ is a phase constant. The latter includes phase shifts introduced by the RF probe positioning, which in turn determines some variations in the location of reference planes, and VNA signal processing adding multiples of $2\pi$. $\eta(k)$ represents the antenna excitation efficiency, proportional to the spatial Fourier transform of the magnetic field produced by the antenna (see Section 3 of the Supporting Information) and set equal to zero for frequencies below the bottom of the SW band. The whole fit function has been convoluted with a gaussian function whose FWHM (200 MHz) is the experimental linewidth of the FMR curve for our CoFeB films.

| $M_s$ [A/m] | $A$ [J/m] | $\alpha$ | $\gamma/2\pi$ [GHz/T] | $w_{CoFeB}$ [μm] | $t_{CoFeB}$ [nm] | $w_{antenna}$ [μm] | $t_{antenna}$ [nm] | $t_{spacer}$ [μm] |
|---|---|---|---|---|---|---|---|---|
| $1.3 \cdot 10^6$ | $18 \cdot 10^{-12}$ | 0.0043 | 28 | 3.6 | 26.3 | 1 | 129 | 70 |

**Table 1.** Micromagnetic and geometry parameters used from the fit of Im($S_{12}$) signals measured on reference CoFeB conduits not surrounded by MFC and SmCo permanent magnets

The fit parameters are listed in Table I, while in Figure 2b we report for each curve the experimental applied field Ha (first number outside the brackets, in mT) and the bias field $H_0$ used to obtain the best fit (number within brackets). The geometrical and material parameters in Table I are in good agreement with SEM measurements and material characterizations carried out by vibrating sample magnetometry and FMR. It should be noted that the effective field (He) used for the derivation of dispersion relations is the sum of $H_0$ and of the demagnetizing field ($H_M$=-6.3 mT) estimated by the micromagnetic simulation of the conduit under action of a transversely applied field which causes its saturation: He = $H_0$+ $H_M$. A good fit is obtained above 20 mT, with $H_0$ = Ha, in agreement with the fact that only above that field a clear DE configuration with $M_0$ perpendicular to the direction of propagation is set in the conduit. Below 20 mT, instead, we observe a different kind of oscillations that can be ascribed to hybrid modes (between well-defined DE and backward volume (BV) modes) arising from a multidomain configuration with average $M_0$ displaying a sizable longitudinal component (See Figure 3). The fit function of Eq. 1 does not apply to this situation and, consequently, no value for the bias field $H_0$ can be retrieved from the fit.

Figure 2c reports the impulse response $s_{12}(t)$ of reference conduits at the same fields considered in Figure 2b. Starting from $S_{12}(f)$ signals taken at constant input power (-10 dBm) using a harmonic frequency grid [δf, 2δf, 3δf, …, Nδf], with δf = 10 MHz and N=$10^4$ frequency points in each sweep, $s_{12}(t)$ is obtained as the Fourier transformation of the Hermitian completed $\tilde{S}_{12}(f)$ function:

$$s_{12}(t > 0) = \mathcal{F}(\tilde{S}_{12}(f)) \qquad (2)$$

Mathematically this represents the response to an impulse corresponding to a normalized *sinc* function with duration (Δt between the first zeros) of $2/f_{max} = 2/(N\delta f) = 50$ ps. For Ha > 20 mT a wavepacket is clearly distinguishable, whose arrival time at the output RF antenna increases from about 0.4 to 0.6 ns with the applied field, as expected because the group velocity for DE SW decreases with the applied field. At 20 mT, additional oscillations are seen at larger arrival time, after the main wavepacket, thus signaling contributions from other modes than the pure DE one. Below 20 mT no clear response is seen, apart from a residual signal close to t = 0 s which can be ascribed to a non-perfect cancellation of the direct electromagnetic coupling between antennas via time-gating (see Section 3 of the Supporting Information).

Overall, the analysis of reference conduits provided a way for checking the consistency of our methodology and calibrate our fitting procedure to retrieve quantitative information from the fit of data taken on devices including MFCs and permanent magnets.

VNA measurements taken on a fully integrated device, with a distance D = 12 μm between the SmCo magnet and the MFC made of Py, are shown in panels 2d,e,f. To ensure a full saturation of the SmCo hard phase and maximize the internal bias field, samples corresponding to panels from 2d to 2i were first magnetized at 5.4 T by a magnetic pulse as described in the methods. Indeed, data taken on samples magnetized at 2 T using a quasi-static field (see Section 4 of the Supporting Information) indicate a reduction of the bias field, especially at large distance D. For small D, instead, a smaller difference between samples magnetized at 2 T or 5.4 T is observed, thus suggesting that the increase of the remanence magnetization is compensated by a loss of gain of MFC which are partially saturated due to the close proximity to the SmCo magnets. For devices with integrated magnets the signal associated to SW is obtained upon subtraction of a reference signal taken at an external field which compensates the internal field produced by the magnets, so that the actual bias field felt by the magnonic conduit is null. From the color map of Im($S_{12}$) reported in figure 2d, corresponding to D = 12 μm, the effect of the bias field produced by the combination of permanent magnets and MFC is clearly visible. The whole map is no longer symmetric with respect to Ha=0 mT and we observe also a shrinking along the Ha axis associated to a smaller distance between the two minima and a higher slope of the iso-phase lines. This is consistent with the presence of a positive bias field $H_0$, arising from the magnets, which can be estimated by multiplying the shift of about 5 mT by the gain of the MFC (G=2.2), thus obtaining $H_0 \approx 11$ mT. In a similar way we can quantitatively interpret the shrinking of the color map along the horizontal direction as due to the presence of the MFC, as the distance between the minima is roughly reduced by a factor 2, in agreement with the estimated value of the MFC gain G=2.2. The analysis of the oscillation of Im($S_{12}$) in the frequency and time-domain are reported in panel 2e and 2f respectively, for some selected values of the applied field Ha (in mT) corresponding to the numbers close to each curve outside brackets. The corresponding bias field $H_0$, from the fit of the imaginary part of $S_{12}(f)$ curves, is instead reported within brackets. Noteworthy, at variance with the case of reference conduits of Figure 2b, here we observe some oscillations even at zero applied field signaling that the system sustains the propagation of SW also without need of an external electromagnet. However, oscillations are localized at low frequency and a reliable fit cannot be performed; we are in the region of one minimum of the color map (Fig. 2d) and a model based on pure DE SW cannot be applied. In the time-domain (Fig. 2f) we see a good propagation of DE wave-packets down to 5 mT of applied fields, while at zero applied field the time response is much broader and due to another kind of band dispersion.

A "textbook-like" DE propagation in the CoFeB conduit is seen in devices with D = 0 µm, where the SmCo magnets are touching the MFC, corresponding to panels g,h,i in Figure 2. The color map of Fig. 2g is shifted by about 9 mT towards negative fields, while the distance between the two minima is similar as the MFC are the same in all devices and their gain is poorly affected by the presence of the non-uniform bias field by the SmCo magnets.

The big difference is that now, starting from high positive fields, the iso-phase lines in the color plot exhibit a monotonic decrease and intercept the zero-field vertical axis at positive frequencies indicating that a well-defined DE configuration is set without application of any external bias field. From the fit of some selected $Im(S_{12})$ curves shown in Figure 2h, the values of the bias field in the region of the conduit flanked by the antenna can be retrieved (values within brackets in mT). For $H_a=0$ (bottom blue curve in panel 2h) we estimated a bias field $H_0=18$ mT, in nice agreement with the shift of the color map by 9 mT considering that the gain of the MFC is G = 2.4±0.3. Furthermore, we see that wavepackets can be seen also in the impulse-response at 0 mT, thus confirming that this device represents a first demonstration of a standalone magnonic device, not requiring any external source of magnetic bias field, fully integrated on silicon.

*2.3 Brillouin Light Scattering investigation*

Spin wave propagation has been investigated by BLS in a device with D = 0 µm, where the SmCo magnets are slightly superposed to the MFCs. The sample underwent a first magnetization up to only 2T in a VSM electromagnet, using a quasi-static process.

The sample's layout (Figure 4a) is the same of the devices studied by broadband spectroscopy (reported in Figure 1a), apart from the positioning of the second antenna (not used in the BLS experiments) which is placed out of the region corresponding to the horizontal arm of the T-shaped MFC. First, we analyzed the spin wave propagation in a well-defined DE configuration, with $M_0$ perpendicular to the spin wave wavevector, under an external magnetic field $\mu_0 H_a=100$ mT along the short axis of the waveguide.

To characterize the spin waves modes excited by the antenna in the CoFeB waveguide, the spin wave intensity was measured as a function of the excitation frequency in the range 6 – 16 GHz, at the center of the region flanked by the horizontal arm of the MFCs. As it can be seen in Fig.4b the BLS spectrum shows an intense peak at about 12.6 GHz. Figure 4c shows a two-dimensional map of the spin wave intensity, recorded at fixed frequency, over an area of about 4.5×10 μm$^2$ with a 250 nm step size. As it can be seen this mode is characterized by an almost uniform spatial profile across the width of the waveguide, as expected for the fundamental mode of a transversally magnetized conduit.

Then we studied spin wave propagation at zero external applied field μ$_0$Ha = 0 mT, following the same methodology. Fig. 4e reports the BLS spectrum of spin waves excited by the antenna in the range between 2 GHz and 8 GHz measured at the center of the region flanked by the horizontal arm of the MFCs. In addition to an intense peak at about 3 GHz, one can observe two modes having a small intensity at about 4.75 GHz and 5.5 GHz. Fig.4e shows two-dimensional maps of the spin wave intensity acquired at fixed frequency for the three observed modes as a function of the distance from the antenna over an area of about 6×11 μm$^2$ with a 250 nm step size (Fig.4f). We find that the most intense mode is characterized by an almost uniform spatial profile across the width of the waveguide, while the two modes exhibit a more complex spatial profile. One can also observe that all the modes tend to localize near the left border of the waveguide. This behavior, which was not observed in the measurements at μ$_0$Ha=100 mT, can be ascribed to a little misalignment of the waveguide with respect to the center of MFCs. This misalignment causes a small inhomogeneity of the bias field H$_0$, which becomes negligible when an intense field is applied.

Comparing the frequency measured for the uniform mode with the dispersion relation of DE modes calculated using the analytical model [19][21] (Fig.5f) we estimated a bias field produced by the magnet of about μ$_0$H$_0$=15 mT. This value is slightly smaller than the 18 mT from VNA measurements in Figure 2g from the same device magnetized at 5T, but in good agreement with the value (16 mT) estimated by VNA upon magnetization up to just 2T (See Section 4 of the Supporting Information), as in case of BLS experiments. This estimation is also consistent with the appearing of the two additional slightly intense modes reflecting some inhomogeneities of the static magnetic configuration. In fact, micromagnetic simulations indicate that a single domain configuration is achieved only when an applied transverse field exceeds 18 mT (Fig. 3).

Finally, it is worth noting that at zero applied field the fundamental mode exhibits a propagation distance similar to that observed for a well-defined DE geometry stabilized by the application of an intense field. In particular, the spin wave decay length has been estimated from the fit of the micro-BLS intensity profile taken in the whole conduit, by using the equation $I(y) = I_1 exp\left(\frac{-2y}{\lambda_D}\right) + I_0$, where $y$ is the position along the waveguide, $I_1$ the SW intensity at the antenna position, and I$_0$ the offset baseline due to noise and detector dark count effects. As shown in Fig. 4(d) and (g), we found that the decay lengths are comparable, assuming the values $\lambda_D = (7.6 \pm 0.2)$ μm and $\lambda_D = (5.6 \pm 0.1)$ μm for =0 mT and =100 mT, respectively. This finding confirms that the bias field $H_0$ provided by the micromagnets is strong enough to allow a good spin wave propagation in the DE configuration even at zero applied field, in agreement with the VNA spectroscopy results.

*2.4 Functional properties of standalone devices*

To assess the potential of our devices we investigated in detail the transmission of SW between the input and output RF antennas in standalone devices (i.e. without any applied external magnetic field), as a function of the distance D between the MFC and the permanent micromagnets. Figure 5a shows Im($S_{12}$) curves vs. frequency taken at zero applied field for D = 0, 1.5, 4, 8, 12 μm. For D from 0 to 8 μm we observe well-defined oscillations with a low frequency edge (corresponding to the FMR limit) which shifts towards lower frequency when increasing D, as expected due to the decrease of the bias field $H_0$. In this range of distances, the experimental curves can be nicely fitted using the model described above for DE spin waves. This is no more possible at D=12 μm, where the bias field is not high enough to overcome the anisotropy field arising from shape anisotropy of the CoFeB conduit and a pure DE configuration cannot be achieved. The bias field $H_0$ in the central region of the conduit for the various values of D, as derived from the analysis of $S_{12}(f)$ signals, is shown in Figure 5f. A sizable tuning of the internally generated bias field from 20.5 to 11 mT is obtained, thus providing an easy way to tune functional properties of this proof-of-concept device, namely the time delay and the phase shift of RF signals between the input and output antennas. Figure 5b shows the impulse-response of each device obtained from the analysis in the time-domain of $S_{12}(f)$ described above. Note that the bias field values obtained from the fit of VNA signals of Fig. 5 are slightly higher than those of Fig. 2. This is since the detailed 2D maps of Fig. 2 were measured one month after curves of Fig.5 and we observed in the meantime a degradation of the SmCo micromagnets which is due to a non-optimized fabrication process used in this proof-of-concept device.

The experimental propagation time of the SW wave packet from the two antennas ($\Delta t_{exp}$), which represents the key performance indicator for a device operated as a time-delay line, varies between 370 and 450 ps when increasing D from 0 to 8 μm. To assess the reliability of this analysis we carried out a theoretical estimation of the propagation time in our conduits. In panel 5d we show the band dispersion of DE modes in our conduits according to the analytical model for SWs in ferromagnetic metallic stripes,[20] for values of applied fields corresponding to the $H_0$ from our analysis of $S_{12}(f)$ in standalone devices at various D (see panel 5f), together with the simulated antenna efficiency. It is quite evident that the frequency band in which we observe good propagation of the spin wave signal is compatible with the low-frequency bound imposed by the FMR and the high-frequency one resulting from the loss of antenna efficiency at high wave vector. On the other hand, the downwards shift of the band dispersion at larger D (smaller bias field $H_0$) is not rigid but it is accompanied by an overall increase of the group velocity which translates into a decrease of the theoretical propagation time ($\Delta t_{th} = v_g \cdot r$) shown in panel 5e. This is qualitatively consistent with the trend of the experimental propagation time reported in Figure 5b. On the other hand, the comparison of the absolute values of the experimental and theoretical propagation times of panels 5b and 5e requires some preliminary considerations. According to the time-domain analysis discussed above, the signals in Figure 5b correspond to the impulse-response to an input pulse with duration of 50 ps, corresponding to an electromagnetic RF input signal made of harmonic components with same amplitude between 0 and 40 GHz. However, this wide-band signal is filtered by the transmission band of the magnonic conduit, which is bounded between the FMR and the highest frequency at which we observe a sizable excitation efficiency of the antenna. This is evident from Figure 5d, were we report with black dashed line the product of the antenna efficiency and the exponential attenuation bounded by the FMR ($\beta(k) = \eta(k) \cdot \exp(-r/L_{att})$) for a bias field $H_0$=20.5 mT (D = 0 μm), which defines the effective bandwidth of the magnonic device. Extending this analysis to all values of D we find that our devices feature a SW transmission band from about 3 GHz and 10 GHz, corresponding to an effective bandwidth $\Delta f_e$ = 7 GHz of the input pulse for the magnonic conduit. This translates into longer effective duration of the magnonic input pulse, on the order of $2/\Delta f_e \approx 280$ ps, in good agreement with the duration of main wavepackets seen in panel 5b taking into account that this is also affected by band dispersion. On the other hand, when we estimate $\Delta t_{th}$ as the product of the group velocity ($v_g$) by the distance between the two antennas (r) we are considering a much narrower wavepacket, with bandwidth $\Delta f_g \ll \Delta f_e$, centered at the frequency where $v_g$ is calculated. According to this analysis we can say that the time response seen in Figure 5b is quite coherent with the theoretical propagation times shown in Figure 5e. In fact, within the effective bandwidth for SW in the magnonic conduit $\Delta t_{th}$ are within 0.3 and 1 ns, corresponding to the time window where we observe some wavepackets in the time-domain signals of panel 5b. Furthermore, as the peak of the product of the antenna efficiency and exponential attenuation is found at about 1 rad/μm, we expect that the more relevant contribution to the wavepackets observed in Figure 5b is arising from packets centered around this value. This also corresponds to the calculated propagation times (Figure 5e) in the order of 0.4 ns, in good agreement with the waveforms of Figure 5b.

Considering the device as a "phase shifter," we aim to evaluate the phase change caused by SW propagation between the antennas. This can be estimated from the phase of the scattering parameter $S_{12}(f)$, which is reported in panel 5c in the 3-7.5 GHz range for D = 0, 1.5, 4, 8 μm using as reference the phase of $S_{12}$ at 3 GHz in the device with D=0 μm. The procedure for the extraction of relative phases from $S_{12}(f)$ measured on different devices is detailed in the methods. As expected for the propagation of DE modes, a linear decrease of the phase vs. frequency was observed. It is important to note that the phase shift at fixed frequency can be tuned by about 100 degrees by changing D and a sizable change of the bias field $H_0$, as already demonstrated.

## 3. DISCUSSION

To the best of our knowledge, the devices reported in this paper represent the first example of standalone magnonic devices with electric input/output, fully integrated on silicon and capable of RF signal processing up to 8 GHz, that do not require external magnetic bias fields. Each of them has a footprint of just 100x150 μm$^2$: this is a record for a self-standing magnonic device and shows the huge potential of magnonics for the integration of RF components, also in comparison with the current technology based in SAW if one considers that the typical footprint of a SAW filter is on the order of 1 mm$^2$. They are fabricated with a planar process suitable for on-silicon wafer scale production at reduced costs, compatible with application in consumer electronics.

On top of that, in this paper we demonstrate that our integrated magnonic devices can be also easily tunable because the bias field $H_0$ produced by the combination of permanent micromagnets and MFC can be varied changing the distance D between them. This does not mean that our devices are reconfigurable in operation, as the distance D is fixed during fabrication. However, our design is compatible with the implementation of real-time reconfigurability according to the general idea of the running EU project MandMEMS, which aims to combine MEMS and magnonics.[22] The MFC and permanent magnets could be mounted on the movable parts of MEMS devices allowing for a continuous tuning of the distance D. In this regard we notice that displacements on the order of 10 μm, corresponding to the range of D values explored in this paper, are easily achievable with state-of-the-art silicon-based MEMS. In this perspective, it is worth summarizing the performances of this first batch of demonstrators to evaluate the achievable range of tunability. From data in Figure 5b our devices could be used as time-delay lines with tunable delays up to 150 ps by varying D in the 0-8 μm range. From another perspective they could implement a phase shifter with tunable phase shift, as shown in Figure 5f, reporting the variation of the phase shift at 6 GHz as a function of the distance D. A tunable relative phase shift of about 120 degrees could already be achieved using the layout of this first batch of devices but larger modulations can be easily envisaged.

The simplest way to amplify the effect of the change of the distance D is to increase the strength of the stray magnetic field produced by the permanent magnets. This can be done in two ways: increasing the remanence magnetization of the SmCo pads and/or increasing the thickness. In fact, the stray field goes linearly with the SmCo thickness for a fixed geometry, so that just doubling the thickness up to 2 μm could allow to achieve a bias fields as high as 40 mT for small D and a phase shift of $2\pi$ when increasing the distance D, according of data taken on reference conduits (see Figure 2b).

Other methods for biasing magnonic conduits and tuning the transmission of SW have been proposed, mainly based on the usage of current lines producing variable bias magnetic fields or, more recently, the implementation of magnetoelectric coupling. It is worth to point out here that our method, based on permanent magnet, is much more "green". There is no energy consumption associated to the generation of the bias field while its modulization could be carried out using a piezo or via capacitive actuations, with low power consumption. For example, generating such 20 mT with a current line would require a current density of about $4\cdot10^{-6}$ A/cm$^2$ flowing in a wire with the same width as the magnonic conduit and 1 µm thick, at 1 µm distance from the conduit itself, which could result in a sizable overheating of the whole device.

Finally, note that the signals reported in this paper have been obtained upon subtraction of the reference signal coming from the direct electromagnetic coupling between antennas. Due to the relatively high damping of CoFeB ($5\times10^{-3}$ in our films) the distance $r$ between antennas ensuring a sizable magnitude of SW signal at the receiver cannot exceed 10 µm. In our design r = 5 µm and the RF signal at the output antenna due to SWs is just 1% of the entire signal. This make quite impractical the direct usage of such devices in real applications, even though we stress here that they would be fully compatible with integration in consumer electronics products. A careful engineering and optimization of the actual performances is needed to improve their technology readiness level and qualify them as RF signal processing integrated device, but this is beyond the scope of this work.

## 4. CONCLUSIONS

In this paper we describe the realization of fully integrated and standalone magnonic devices on a silicon substrate, based on the combination of CoFeB conduits and an assembly of magnetic flux concentrators and permanent magnets to internally generate the bias field for stabilizing a DE configuration for SW propagation. The devices have a total footprint of 100x150 µm$^2$ and feature an input and output antenna for RF signals processing in the 3-8 GHz range. We demonstrated that our proof-of-concept devices can operate without any external source of magnetic field, implementing some basic functionalities like that of a tunable time-delay line or phase shifter. The tuning of the time-delay (up to 150 ps) or phase shift (up to 120 deg at 6 GHz) is achieved by setting the distance D between the flux concentrators and the permanent magnets in the 0-8 µm range during fabrication. Future developments of this technology involve the realization of reconfigurable devices, by mounting the permanent magnets on the movable part of a MEMS to vary the distance D and thus the bias field.

## 5. EXPERIMENTAL SECTION

*Device fabrication.* The whole fabrication process is made of four main steps. *1)* 1 μm thick SmCo permanent magnets (100x40 μm footprint) embedded in a Si(001) coupon (20x20 mm) are realized by sputtering deposition in a trench previously defined by reactive ion etching, with a mesoporous silica layer used for subsequent lift off. Upon lift-off an annealing at 650°C for 1 h has been performed to promote the formation of the magnetic hard phase. 2) 1 μm thick MFC made of a $(Py(80)/Cr(5))_{12}$ multilayer (thickness in nm) are defined by optical lithography, sputtering and lift-off. 3) CoFeB waveguides are fabricated on the untouched silicon substrate flanked by the assembly of MFC and permanent magnets by optical lithography, magnetron sputtering and lift-off. 4) Finally, upon deposition of 70 nm of silica, RF antennas are realized by optical lithography, thermal evaporation and lift off of 100 nm of gold with 10 nm of Ti as adhesion layer. Coupons are then cut into 10x10 mm samples containing arrays of devices with different layouts, and subsequently magnetized in a uniform field of 2 T provided by a conventional electromagnet available in our vibrating sample magnetometer setup. The samples used for the electrical characterization by VNA and reported in the main text have been magnetized with a pulse of 5.4 T provided by a magnetizer (Model *i Mag MicroMag* from Laboratorio Elettrofisico Engineering S.r.l.), ensuring a full saturation of the hardest $SmCo_5$ phases typically found in our SmCo films.

*Micro-MOKE.* A home-made microMOKE apparatus was employed in the longitudinal configuration, which involves applying a field parallel to the sample surface while simultaneously recording the reflected signal at a non-normal angle to the surface to maximize the response associated with the in-plane magnetization component. White p-polarized light was used for illumination, while reflected light was analyzed via a polarizer in an s-configuration to obtain the signal related to Kerr rotation, prior to image acquisition using a high-sensitivity and high-resolution charge-coupled device (CCD) camera. By analyzing the variation in magnetic contrast of a selected area of interest, local magnetization curves with micrometric resolution can be extracted (Figure 1c-d). The images in Figure 1f-i were acquired using a 50x objective, enabling the capture of regions of approximately 150x400 μm with a spatial resolution of about 500 nm.

*VNA spectroscopy.* Broadband spectroscopy experiments have been carried out using a home-made RF probe station featuring a quadrupolar vectorial electromagnet suitable to produce in-plane fields as high as 200 mT and a 10 MHz - 43 GHz, 4 ports Vector Network Analyzer (R&S ZNA43).

*Brillouin Light Scattering.* Micro-BLS measurements were performed by focusing a single-mode solid-state laser (operating at a spectral line of 532 nm) at normal incidence onto the sample using an objective with numerical aperture of 0.75, giving a spatial resolution of about 250 nm. The inelastically scattered light was analyzed by means of a (3+3)-pass tandem Fabry-Perot interferometer. A nanopositioning stage allowed us to position the sample with a precision down to 10 nm on all three axes, and to perform spatial resolved scans moving the sample with respect to the objective. A spatially uniform magnetic field $\mu_0 H_A = 100$ mT, provided by an electromagnet, was applied in the sample plane along the short axis of the waveguide. A DC/AC electrical probe station ranging from DC up to 20 GHz was used for spin-wave excitation. The microwave power was set +0 dBm on the RF generator output.

*Simulations.*

Magnetic field simulations were performed using COMSOL Multiphysics 6.1 to optimize the system geometry. The material properties were incorporated using the Jiles-Atherton model to reproduce the experimentally measured M(H) hysteresis loops of the $(NiFe(80)/Cr(5))_{12}$ multistack and the CoFeB waveguide. The SmCo micromagnets were assumed to be fully saturated along the desired direction, with a magnetization value corresponding to the remanent magnetization measured via VSM. This assumption is justified by the fact that both the stray field and the applied field remain significantly lower than the coercive field required to alter the magnetization. The shape and dimensions of the field concentrators were iteratively modified and optimized to simultaneously enhance the magnetic field concentrated at the waveguide positions and improve sensitivity to variations in the micromagnets' positions, thereby increasing tunability. More details on the simulations can be found in the Supporting Information. The simulations of the equilibrium micromagnetic configuration and SW dispersion for different applied transverse magnetic field have been carried out using the mumax3 software.

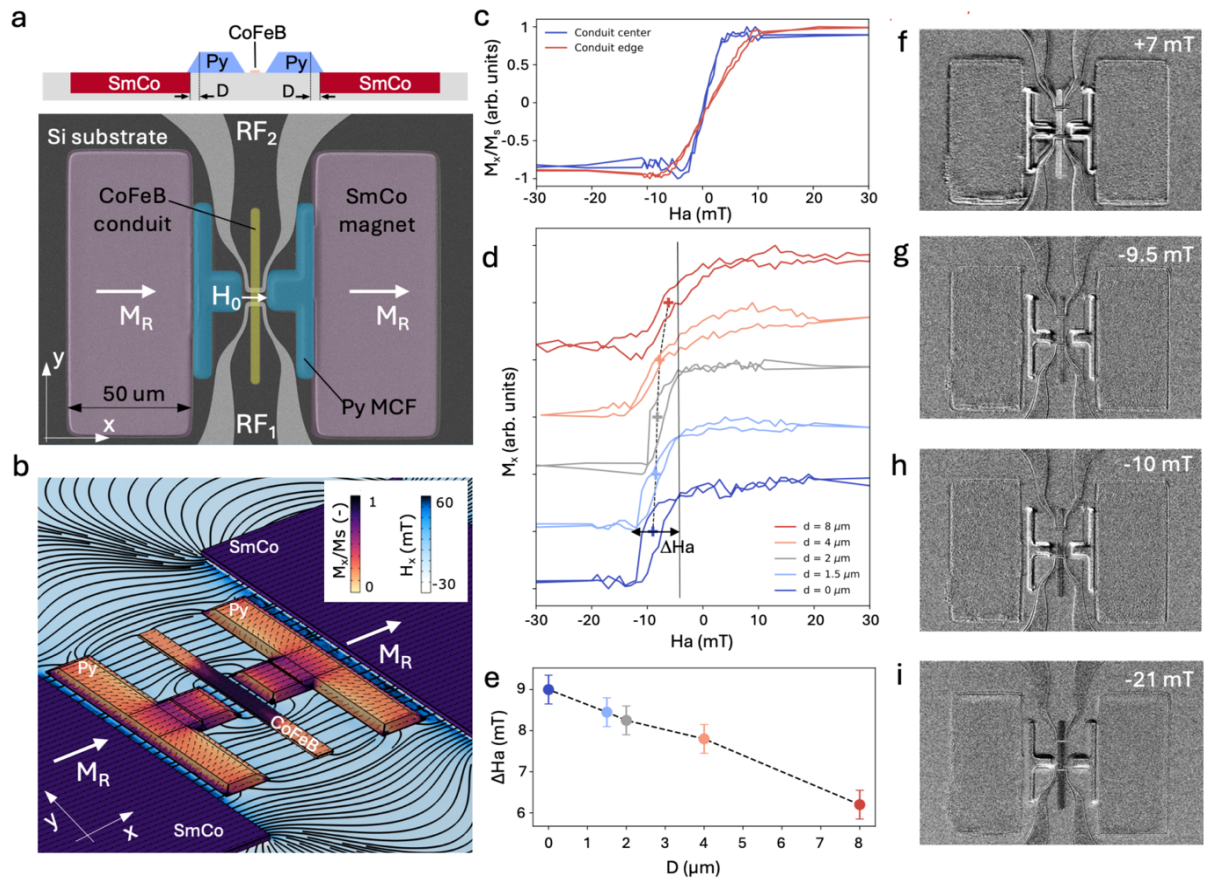

**Figure 1.** a) False-color SEM image and schematic cross section of a device. b) COMSOL Multiphysics simulation of the stray field H produced by the assembly of SmCo permanent magnets and MFC made of Permalloy (field lines and blue color scale for the $H_x$ component), combined with the simulation of the micromagnetic configuration of the CoFeB conduit (arrows and Brown color scale for the $M_x$ component). c) Mx vs Hx loops measured at the center (within the poles of the MFC) and edge of the CoFeB conduit, prior to the initial magnetization at 2T of the SmCo magnets. d) Loops measured at the center of the conduit after magnetization of the SmCo micromagnets at 2T, in devices with different distance D between the MFC and the SmCo magnets. e) Shift $\Delta Ha$ of the loops reported in panel d for various values of D. f,g,h,i) Snapshots from a video (available as Supplementary Material) taken with a microMOKE apparatus during a whole loop, at characteristic applied fields Ha= +7, -9.5, -10, -21 mT, respectively.

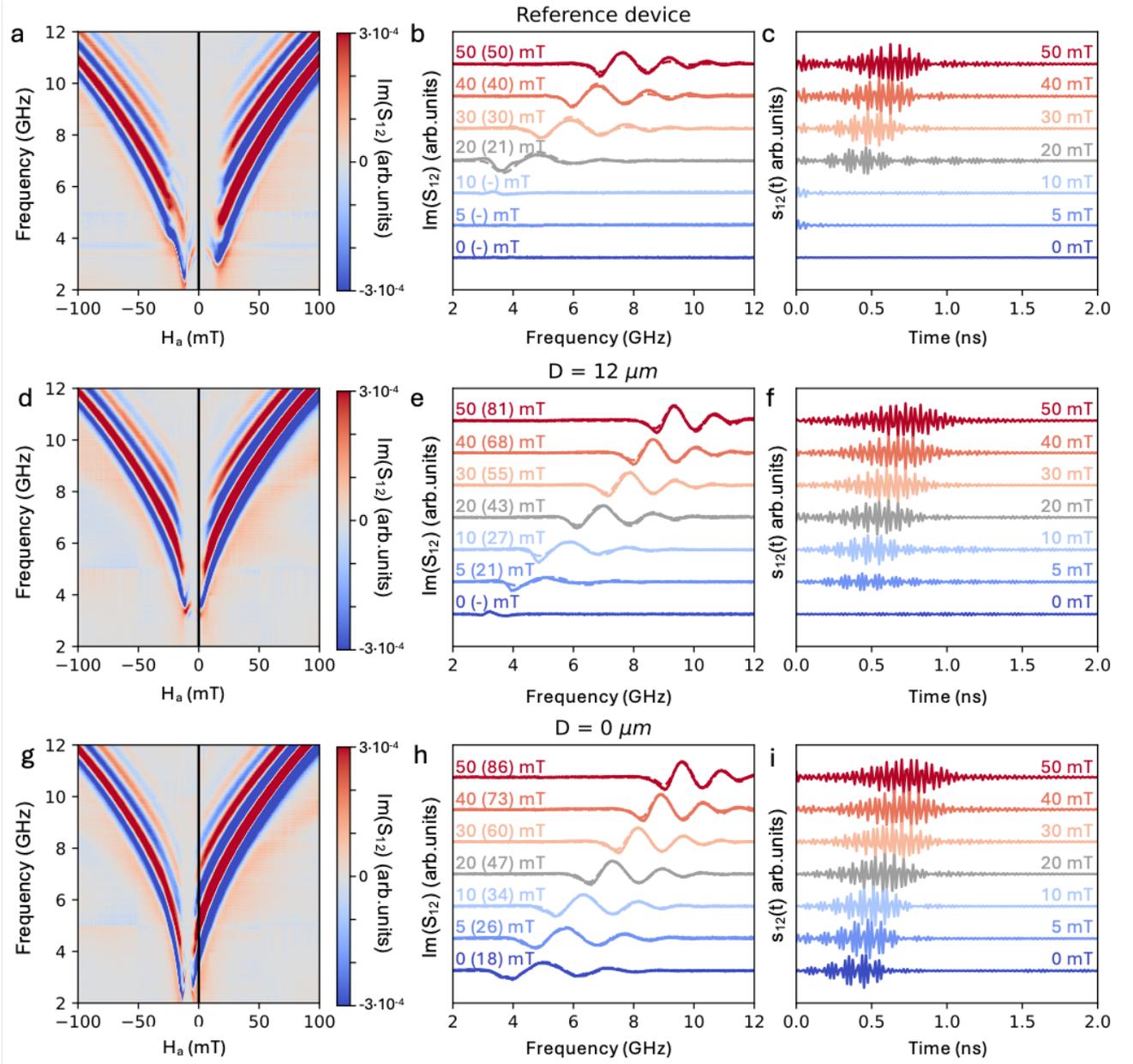

**Figure 2.** Panels a,b,c report VNA measurements on reference CoFeB conduits without MFC and permanent magnets. a) 2D color map of the oscillations related to the imaginary part of $S_{12}$ vs frequency for various applied fields $H_a$. b) Experimental $S_{12}(f)$ curves (continuous lines) for selected values of $H_a$ (numbers in mT outside the brackets close to each curve), together with the result of the fit using equation 1 (dashed lines) from which the actual values of the effective applied bias field (numbers in brackets) are retrieved. c) Impulse response $s_{12}(t)$ obtained from the time-domain analysis of the $S_{12}(f)$ scattering parameters corresponding to the curves in figure 2b. Panels d,e,f report the same analysis as in a,b,c for devices with a distance between MFC and permanent magnets $D = 12$ μm. Panels g,h,i refer to a device with $D = 0$ μm.

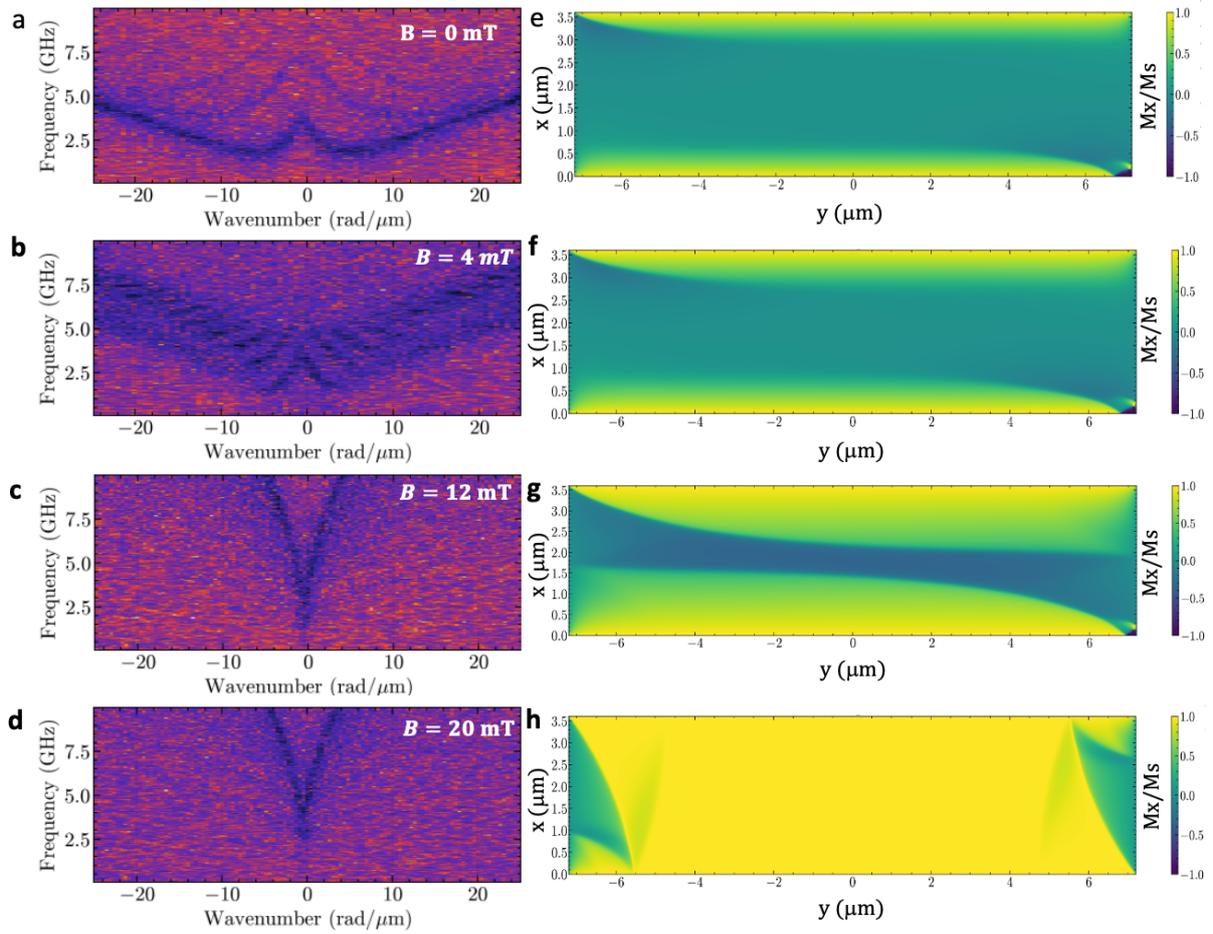

**Figure 3**. Panels a,b,c,d: band dispersion simulated with mumax3 for reference CoFeB conduits at different applied field B = μ₀Ha. Panels e,f,g,h: corresponding static micromagnetic configurations.

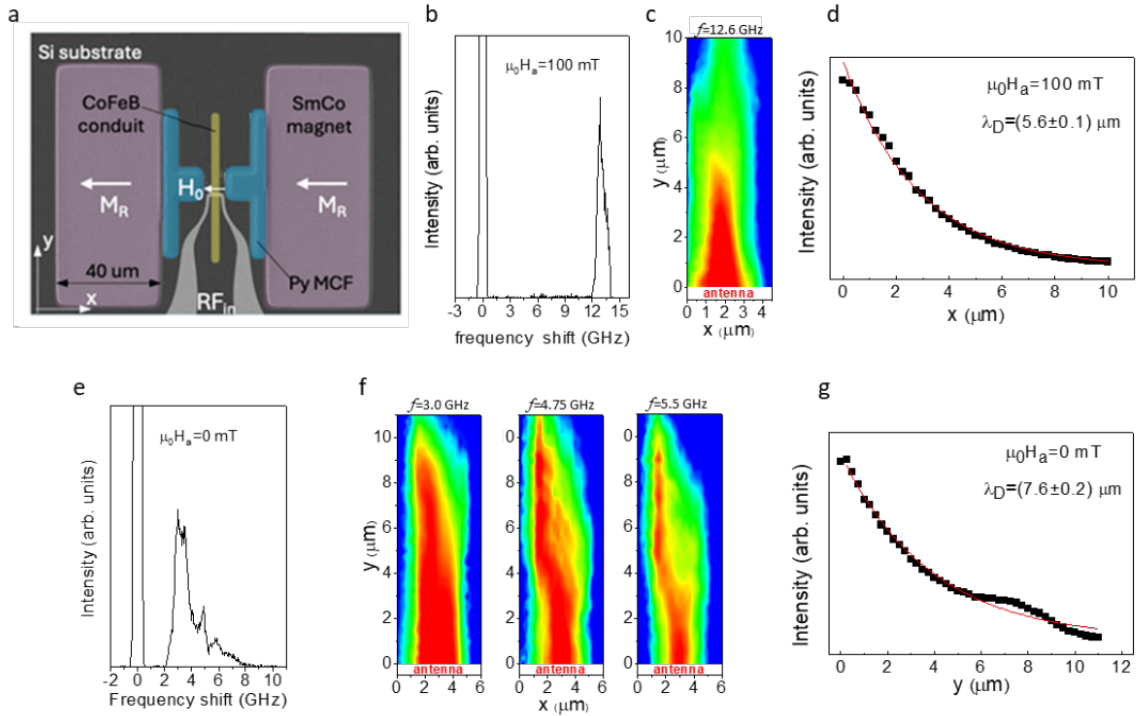

**Figure 4:** a) False-color SEM image of the investigated device. b) and e) Spin wave spectra recorded by micro-BLS at $\mu_0H_a$ =100 mT and $\mu_0H_a$ = 0 mT, respectively. c)Two-dimensional map of the BLS intensity recorded at a frequency of 12.6 GHz excited by the inductive antenna. An external field $\mu_0H_a$ =100 mT was applied along the short axis of the conduit. d) and g) Spin-wave intensity (linear scale) as a function of the propagation distance y from the antenna for $\mu_0H_a$ =100 mT (d) and $\mu_0H_a$ = 0 mT (g), respectively. Black points correspond to experimental data, whereas red lines show the exponential fit considering $\lambda_D$ = (7.6 ± 0.2) μm. e) Two-dimensional map of the BLS intensity recorded for the modes at frequencies of 3GHz, 4.75 GHz and 5.5 GHz, respectively at $\mu_0H_a$ = 0 mT.

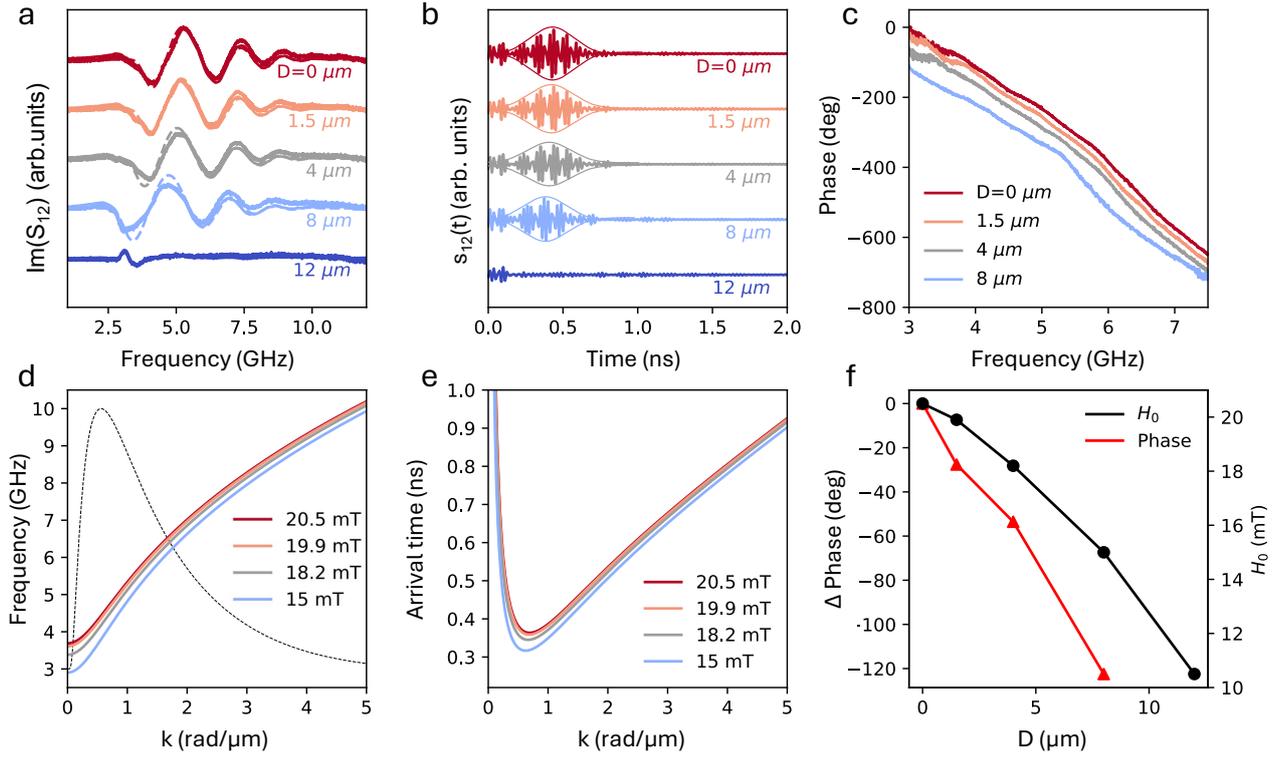

**Figure 5** a) Im($S_{12}$) curves measured on devices with different values of D at Ha = 0 mT (when using them as standalone devices without any applied magnetic field). b) Impulse response of the same standalone devices from the time-domain analysis of $S_{12}(f)$ scattering parameters. c) Relative phase of $S_{12}(f)$ for devices with various D values. d) Continuous lines: calculated band dispersions of DE spin waves for bias field values $H_0$ reported in the legend and corresponding to D = 0, 1.5, 4, 8 μm. Dashed line: β(k) function defining the bandwidth of the magnonic device for the case of D = 0 mm. e) Estimated propagation time of wavepackets from the two antennas according to the formula $\Delta t_{th}$ = $v_g \cdot r$, where $v_g$ is the group velocity calculated from the band dispersions of panel 5d and r is the distance between the input and output antennas. f) Right scale: bias field $H_0$ estimated from the fit of curves in panel 5a. Left scale: phase shift introduced in magnonic devices with different D at 6 GHz.

Acknowledgements

All the authors acknowledge funds from the European Union via the Horizon Europe project "MandMEMS", grant 101070536. M. M., R. S. and S.T. acknowledge financial support from NextGenerationEU National Innovation Ecosystem grant ECS00000041–VITALITY (CUP B43C22000470005 and CUP J97G22000170005), under the Italian Ministry of University and Research (MUR). R. B. acknowledges funding from NextGenerationEU, PNRR MUR – M4C2 – Investimento 3.1, project IR_0000015 – "Nano Foundries and Fine Analysis – Digital Infrastructure (NFFA–DI)", CUP B53C22004310006. We thank M. Ferretto A. Del Prete and S. Tizianel from Laboratorio Elettrofisico for providing access to the magnetizer used for magnetization of SmCo magnets at 5.4 T.

This work has been partially performed at Polifab, the micro and nanofabrication facility of Politecnico di Milano.

# Supporting Information

## Self-biased integrated magnonic devices

M. Cocconcelli,* F. Maspero, A. Micelli, A. Toniato, A. Del Giacco, N. Pellizzi, A. E. Plaza, A. Cattoni, M. Madami, R. Silvani, A. A. Hamadeh, C. Adelmann, P. Pirro, S. Tacchi, F. Ciubotaru, R. Bertacco*

**Section 1: Static simulations of the bias field from the micromagnets**

Simulations were performed using COMSOL Multiphysics 6.1 to design and optimize the magnetic field configuration for the system under study.

The SmCo micromagnets are modeled as uniformly magnetized along the x-axis, with a magnetization value corresponding to their remanent magnetization, which was estimated experimentally via vibrating sample magnetometry. The magnets are designed to have an elongation along the direction orthogonal to the magnetization direction, to have an aspect ratio that maximizes the intensity of the x-component of the field and to ensure uniformity in the y-direction. The magnets are also designed to be as thick as possible within fabrication limits, which constrain the thickness to 1 micrometer.

To enhance field uniformity, two symmetrically placed magnets magnetized along the x-axis were employed. The intensity and uniformity of the magnetic field in the region between the magnets were found to strongly depend on their spacing, d. While reducing d increases the field intensity, it also compromises uniformity.

Magnetic flux concentrators (MFCs) were introduced to enhance both the field intensity and uniformity in the region of interest, where the magnonic conduit is located. Unlike configurations with single or paired magnets, the introduction of MFCs ensures that varying the spacing between the permanent magnet (PM) and the MFC alters only the field intensity without significantly affecting the field-line distribution.

Existing literature extensively examines the effects of MFC geometry on field amplification [Xiaoming Zhang et al, AIP Advances 1 December 2018; 8 (12): 125222]. However, previous studies primarily assume uniform external fields, whereas our work focuses on fields generated by permanent micromagnets operating at remanence. Therefore, the MFC geometry was systematically optimized via simulations to maximize the amplification of the micromagnets' fields. Several shapes, including T-shaped, triangular, half-circle, wide-bar, and bar configurations, were tested. Among these, T-shaped concentrators showed superior performance for concentrating the non-uniform field produced by rectangular SmCo micromagnets and were further refined to maximize field concentration.

The objectives of this study are twofold: first, to optimize the magnets for maximizing field strength, and second, to optimize the MFC geometry for enhancing field amplification. Due to the interdependence of these components, a recursive design approach was employed. For instance, the MFC length dictated the minimum allowable spacing between the magnets, directly influencing the resulting field.

All relevant design parameters were systematically varied, starting from an initial configuration dictated by fabrication limitations, leading to the optimized values reported in Table S1 and Figure S1 a. The MFCs were designed to be as thick as possible with the thickness of Cr/MoNiFe multilayer limited to 1 μm. The MFCs consisted of alternating layers of Cr (5 nm) and MoNiFe (80 nm), a design choice intended to enhance magnetic permeability compared to bulk MoNiFe. This approach maximizes the gain of the concentrators.

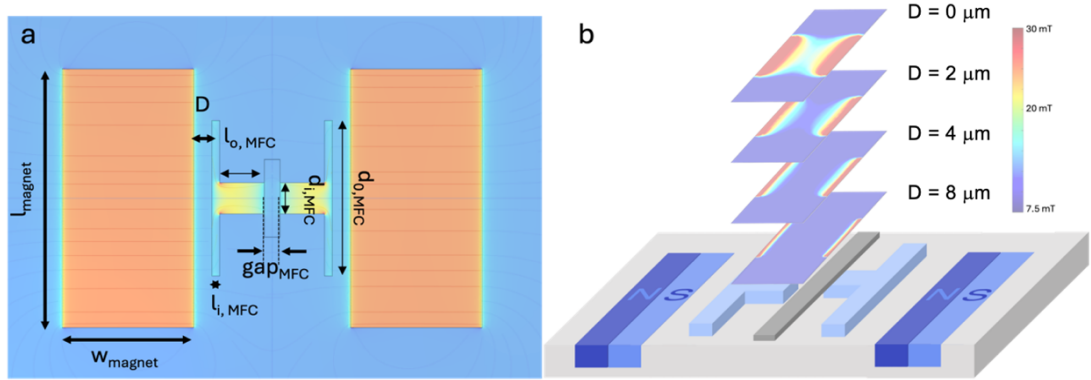

**Figure S1:** a) Layout (top view) of the system; b) Field concentrated by the MFCs in between their poles at varying distances (D) between the magnet and the MFC.

| Magnet length | Magnet width | $\mu_{r,py}$ | $M_r$ | Magnet thickness | $d_{i,MFC}$ | $d_{o,MFC}$ | $l_{tot,MFC}$ | $gap_{MFC}$ | MFC-magnet distance | Magnet z alignment with MFC |
|---|---|---|---|---|---|---|---|---|---|---|
| 100 μm | 50 μm | 1000 | 1 T | 1 μm | 12 μm | 60 μm | 20 μm | 6 μm | 7 μm | Same plane |

**Table S1**: Optimized parameters coming from COMSOL simulations and adopted for the design of devices investigate dint he present paper.

The MFCs were positioned to minimize the gap between them, thereby increasing the field intensity while ensuring that their poles extended beyond the magnonic conduit width by a sufficient margin to accommodate lithographic resolution constraints during fabrication. The relative positioning between the MFCs and permanent magnets was also analyzed, as shown in Figure S1b. Reducing the gap between the magnets and the MFCs significantly enhanced the concentrated field strength. The simulations revealed that adjusting this gap could modulate the magnetic field by approximately 30 %, suggesting the potential for dynamic field tuning using MEMS actuators.

The T-shaped concentrator geometry and dimensions were selected to maximize the magnetic field over a region large enough to facilitate spin-wave excitation, propagation, and detection. The design also accounted for the minimum field of 18 mT required to fully bias a 3-micrometer-wide magnonic conduit made of CoFeB and set a DE configuration. The final optimized geometry, shown in Figure S1a, is expected to produce a field of 24 mT at the waveguide location, as seen in the field profiles shown in Figure S1b. The optimized geometrical parameters are reported in Table S1.

**Section 2: Comparison between MOKE loops taken at the center and edge of the conduits**

Micro-Optical Kerr-Effect (MOKE) measurements were performed on the samples after the SmCo micromagnets were magnetized with a field of 2 T. Hysteresis loops of the CoFeB conduit were measured both at the center of the waveguide (between the poles of the magnetic flux concentrators (MFCs), where the field concentration is strongest), and at the edges of the conduit, where the concentration effect is weaker. The measurements were carried out for various distances between the magnets and the MFCs.

As shown in Figure S2, when the magnets are positioned close to the MFCs (D = 0), the hysteresis loop at the center of the conduit shows a significant shift, with a remanence value at zero external field comparable to the value at saturation. In contrast, the loop measured at the edge of the conduit displays a smaller overall shift and distinct jumps in the hysteresis curve. These jumps are attributed to Barkhausen noise, caused by the formation and movement of magnetic domains along the CoFeB conduit during magnetization reversal, as shown in Figure 1 in the main text. These domains propagate through the conduit and become pinned at regions where the field intensity varies, such as at the edges of the MFCs.

As the distance between the magnets and the MFCs increases, the ability of the MFCs to concentrate the magnetic field diminishes. Consequently, the loops show a smaller shift toward negative fields, and the hysteresis loops at the edge and center of the conduit become more similar. This is because the difference in the internal field between these two regions decreases as the distance increases.

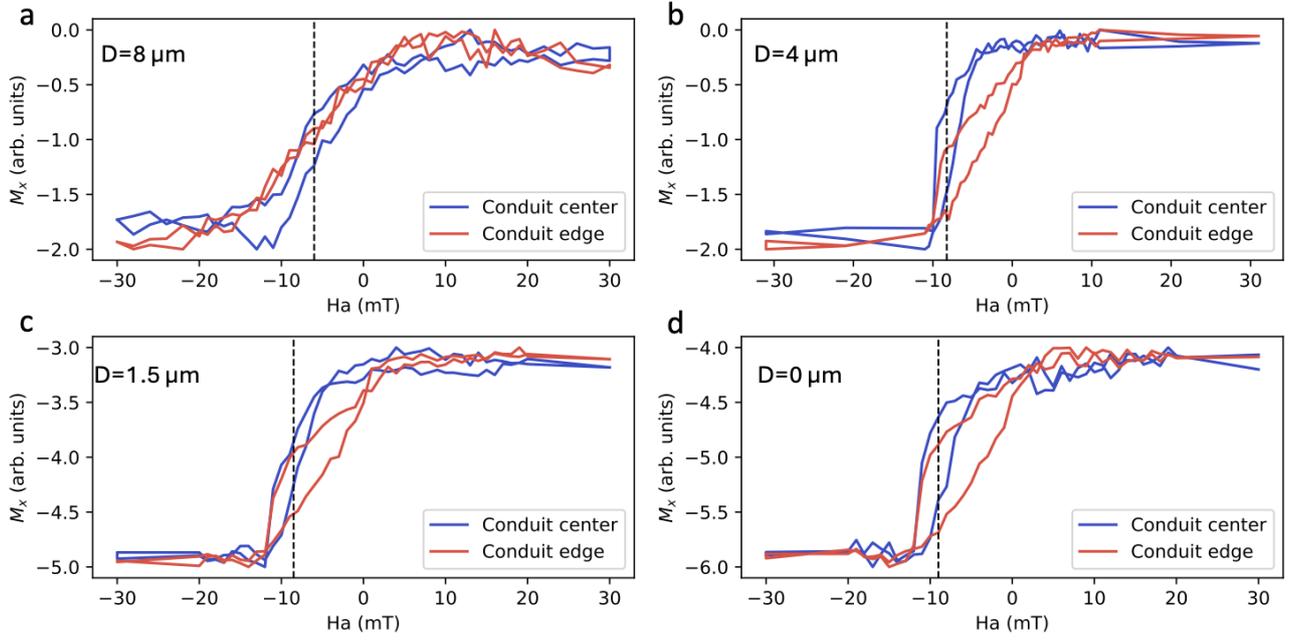

**Figure S2:** MOKE hysteresis loops measured at the center and at the edge of the conduit after magnetization of the SmCo micromagnets at 2T, in devices with different distance D between the MFC and the SmCo magnets.

**Section 3: Data analysis of VNA spectra**

*VNA measurements and time gating*

All-electrical measurements were performed on the samples using a Vector Network Analyzer (VNA), enabling the retrieval of the scattering matrix (S). We focused on the $S_{12}$ parameter, which represents the complex transmission signal emitted from antenna $RF_2$ and received at antenna $RF_1$ (see Figure 1 in the main text).

The transmitted signal comprises both the spin-wave component and a background signal arising from electromagnetic coupling between the two antennas. Due to the small dimensions of the magnonic conduit and the intrinsic damping of CoFeB, the spin-wave contribution constitutes only a minor fraction of the total signal. To isolate the spin-wave signal, a reference subtraction method was employed to remove the electromagnetic noise. For the devices studied, the reference signal was obtained by applying an external magnetic field that counteracted the on-chip field generated by the ensemble of magnetic materials. Under these conditions, the reference signal contained no spin-wave contribution.

However, even after reference subtraction, residual noise can persist due to temporal drift in the background or variations in the direct electromagnetic coupling as a function of the applied external field due to small displacements of the RF probes. Thus, to further reduce the noise, a time-gating procedure was applied. To enable time gating, measurements were conducted using a harmonic frequency comb (f, 2f, 3f, …), starting at 4 MHz and extending up to the maximum frequency achievable with the setup (40 GHz) for a total of 10000 points. The signal was then expanded to negative frequencies by filling in the complex conjugates of the positive frequency measurements and interpolating the value at 0 Hz from neighboring points. An inverse Fast Fourier Transform (FFT) was then performed to convert the frequency-domain signal into the time domain. The time-domain resolution was determined by the maximum frequency measured.

In the time domain, distinct signal packets corresponding to the spin-wave signal could be identified. A Hamming window function was applied to isolate the spin-wave signal, after which an FFT was performed to convert the signal back to the frequency domain. This procedure effectively filtered out noise, yielding a clean spin-wave signal, as shown in Figure S3.

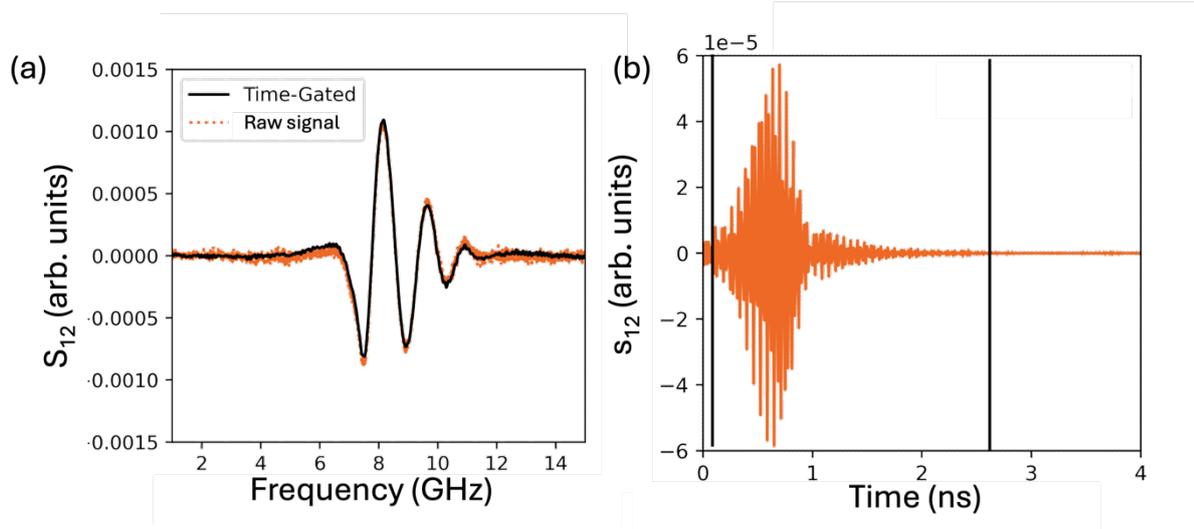

**Figure S3**: Signal from a reference device (CoFeB conduit without other on-chip magnetic elements) under an external magnetic field of 30 mT, shown in (a) the frequency domain and (b) the time domain. In panel (b), the black lines mark the boundaries of the window used for time gating. The gated signal corresponds to the black curve in panel (a).

This procedure is particularly effective when multiple modes are excited, as these modes may overlap within the same frequency band, making it challenging to distinguish between them. However, owing to their differing group velocities, the modes manifest as distinct packets in the time domain. By selectively applying time gating to these packets, it is possible to isolate the contributions of the individual modes in the frequency domain.

*Model for fitting Im($S_{12}$)*

To model the output of the Vector Network Analyzer (VNA), specifically the imaginary part of $S_{12}$, a one-dimensional model can be used, incorporating the dispersion relation and the antenna coupling efficiency [F. Ciubotaru et al, Appl. Phys. Lett. 109, 012403 (2016)]. The resulting expression is given by:

$$\text{Im}(S_{12}(\omega)) = S \cdot \cos(k(\omega)r+\varphi) \cdot \exp(-r/L_{att}) \cdot \eta(\omega) \quad (1)$$

Where S is a coefficient, $k(\omega)$ represents the dispersion relation, r is the center-to-center distance between the antennas, $\varphi$ is an additional phase introduced by the VNA, $\eta(\omega)$ is the antenna efficiency and $L_{att}$ is the attenuation length of the spin wave.

The efficiency of the inductive antenna, $\eta(k)$, is calculated by taking the Fourier transform of the magnetic field generated by the current flowing through the antenna. This field, evaluated at the position of the magnonic waveguide (separated from the antenna by a 70 nm $SiO_2$ interlayer), is computed considering the finite width and thickness of the antenna, which are 1 μm and 129 nm, respectively. The in-plane and out-of-plane components of the magnetic field are given by [R. Buschauer, Phys. Teach. 52, 413–414 (2014)]:

$$H_x(x,z) = -\frac{I}{8\pi ab}\left[(a-x)\left[\frac{1}{2}\ln\left(\frac{(b-z)^2+(a-x)^2}{(a-x)^2+(-b-z)^2}\right) + \frac{b-z}{a-x}\cdot atan\left(\frac{a-x}{b-z}\right) - \frac{-b-z}{a-x}\cdot atan\left(\frac{a-x}{-b-z}\right)\right] - \right.$$

$$\left.(-a-x)\left[\frac{1}{2}\ln\left(\frac{(-a-x)^2+(b-z)^2}{(-a-x)^2+(-b-z)^2}\right) + \frac{b-z}{-a-x}\cdot atan\left(\frac{-a-x}{b-z}\right) - \frac{-b-z}{-a-x}\cdot atan\left(\frac{-a-x}{-b-z}\right)\right]\right]$$

$$H_z(x,z) = \frac{I}{8\pi ab}\left[(b-x)\left[\frac{1}{2}\ln\left(\frac{(b-z)^2+(a-x)^2}{(-a-x)^2+(b-z)^2}\right) + \frac{a-x}{b-z}\cdot atan\left(\frac{b-z}{a-x}\right) - \frac{-a-x}{b-z}\cdot atan\left(\frac{b-z}{-a-x}\right)\right] - \right.$$

$$\left.(-b-z)\left[\frac{1}{2}\ln\left(\frac{(a-x)^2+(-b-z)^2}{(-a-x)^2+(-b-z)^2}\right) + \frac{a-x}{-b-z}\cdot atan\left(\frac{-b-z}{a-x}\right) - \frac{-a-x}{-b-z}\cdot atan\left(\frac{-b-z}{-a-x}\right)\right]\right]$$

Here, a and b represent the half-width and half-thickness of the antenna, respectively, while *x*, *z* are the transverse and out-of-plane coordinates of a reference system centered at the midpoint of the antenna. The Fourier transform of the field produced by the antenna provides $\eta(k)$. This function, initially defined over the k-space determined by the spatial extent *x* used in the calculation, is mapped through interpolation onto the $k(\omega)$ values corresponding to the dispersion relation. This mapping enables the evaluation of the antenna's excitation efficiency at each frequency of interest, $\eta(\omega)$.

The dispersion relation $\omega(k)$ for spin waves propagating in a transversally magnetized waveguide and its relaxation time, $\tau(k)$, are determined using a modified Kalinikos and Slavin model [M. P. Kostylev, G. Gubbiotti, J.-G. Hu, G. Carlotti, T. Ono, and R. L. Stamps, "Dipole-exchange propagating spin-wave modes in metallic ferromagnetic stripes", Phys. Rev. B 76, 054422 (2007)], using the material and geometrical parameters of the CoFeB waveguide reported in the main text and on the actual value of the external field.

The attenuation of spin waves is governed by the ratio between their mean free path and the separation distance between the antennas. The mean free path is calculated as the product of the relaxation time $\tau(k)$ and the group velocity $v_g(k)$. The group velocity is obtained from the derivative of the dispersion relation curve.

By combining these functions, the oscillatory behavior of the imaginary part of $S_{12}$, $Im(S_{12}(f))$, is reconstructed. To account for experimental broadening, the resulting signal is convoluted with a Gaussian function characterized by a full-width at half-maximum of 200 MHz, which corresponds to the experimental linewidth of the ferromagnetic resonance for the CoFeB films used in the study.

The model relies on several parameters, including the geometrical dimensions of the waveguide and antennas, the value of the applied external field, and the material properties of the CoFeB, as summarized in Table 1 of the main text.

To validate the accuracy of the model for our system, we applied it to reference samples. These consisted of CoFeB waveguides with known dimensions (without any additional magnetic materials nearby), placed in a uniform, externally applied field, and excited by stripline antennas identical to those used in the devices under study.

Fitting was performed using the model by fixing the external field and fine-tuning the magnetic and geometric parameters. Initial values for magnetic properties, such as saturation magnetization (Ms), damping constant ($\alpha$), and exchange stiffness (A), were based on ferromagnetic resonance measurements conducted on continuous films of the same material. The initial geometric dimensions were obtained from scanning electron microcsopy. The best-fit parameters showed only minor deviations from the initial estimates, demonstrating both the accuracy of the model and its sensitivity to small variations in the magnetic and geometrical properties.

After confirming the reliability of the model, we applied it to fit the transmission signal recorded using the VNA for the devices under test. The goal was to estimate the internal magnetic field within the waveguide. For these fits, we kept the magnetic material parameters and all the other geometrical parameters fixed at the values emerging from the fit of the reference devices while considering the external field as free parameter for the fit. This approach produced accurate fits, as illustrated in Figure 2 of the main text, and provided the way to estimate the internal magnetic field considered in the modified Kalinikos and Slavin model, considering the finite width of conduit. From the internal bias field, the equivalent bias field produced by the micromagnets was determined by subtracting (adding in magnitude) the demagnetizing field, which was calculated using micromagnetic simulations and found to be 6 mT.

*Phase signal from VNA measurements*

The VNA allows for the extraction of the phase accumulated by spin waves as they propagate between two antennas. This is achieved by analyzing the phase component of the complex $S_{12}$ matrix element.

However, the phase signal may exhibit 360° discontinuities due to mathematical issues. Additionally, noise can introduce irregularities, resulting in spikes that complicate accurate signal interpretation.

In general, the phase of $S_{12}(f)$ corresponding to spin waves is expected to decrease monotonically when f increases, with a slope depending on the antenna separation and the dispersion relation, which allows for an easy differentiation from the background signal.

To accurately determine the accumulated phase for the various devices, the following procedure was implemented:

1. Time gating was applied to the $S_{12}$ matrix element to reduce noise and isolate the spin-wave signal.
2. An offset correction was performed to eliminate 360 deg phase jumps, ensuring that the onset of the spin-wave signal started at 0 deg.

To further refine the estimation of phase variation and validate the accuracy of the procedure, the oscillatory behavior of the fitting function used to model the acquired signals was analyzed. The phase variation was then determined by calculating the phase differences between the curves corresponding to different devices, as shown in Figure S5. The phase difference among signals from devices with different D was then used to find the realative shift of the experimental curves $S_{12}(f)$ in order to compare them as in Figure 5c of the main text.

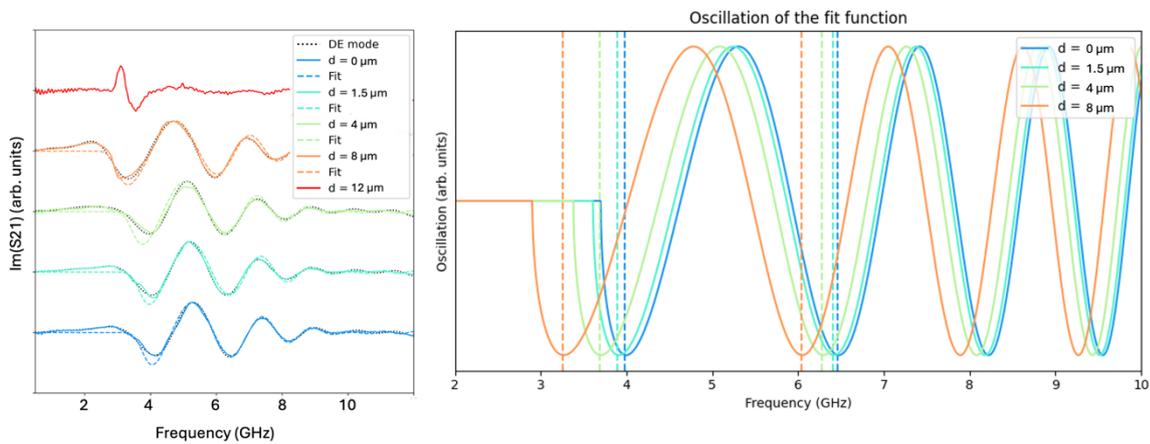

**Figure S4**: (a) Signal in the frequency domain (continuous line), gated signal corresponding to the main mode (black dotted lines) and fit function (colored dotted lines). (b) oscillating part of the fit signal, undamped, used for the estimation of the phase variation.

**Section 4: VNA analysis on devices with SmCo magnets magnetized at 2T**

VNA measurements were carried out upon magnetization of the SmCo micromagnets. Initially, the micromagnets were magnetized using a 2 T field (the maximum filed available in the laboratories of Politecnico di Milano) while only in a second step VNA measurements where refined upon magnetization at higher field (5.4 T) using a magnetizer kindly provided by Laboratorio Elettrofisico.

While in Figure 2 and 5 of the main text we show VNA measurements after magnetization with 5T, here (Figure S4) we report data previously taken upon magnetization at 2 T, showing than in the latter case the bias field from the micromagnets was slightly lower. By applying the fitting procedure described earlier, also this case it was possible to estimate the field concentrated by the MFCs in the absence of an external field. For the case where the distance between the magnets and the MFCs was zero micrometers (D = 0 um), the bias field was found to be approximately 16 mT, in fair agreement with the shift of the symmetry axis of the 2D map of about 7.4 mT if one consider the gain of the MFC (G ~ 2.4, see main text). Notice the this bias field is slightly lower than the value of 18 mT of Figure 2 (1 month after magnetization at 5T, after some physical degradation of the SmCo magnets) and definitely lower than the 20.5 mT measured just a few days after magnetization at 5.4 T. Devices with greater distances between the magnets and the MFCs showed much weaker signals at 0 mT of applied field. The shapes of these signals suggest that the conduit configuration was not a pure DE one, with the magnetization vector displaying a longitudinal component which could account for the BV character of the $S_{12}(f)$ signals measured in this case. This is another indication that the magnetization at 5.4 T instead of 2T was effective in improving the bias field provided by the micromagnets.

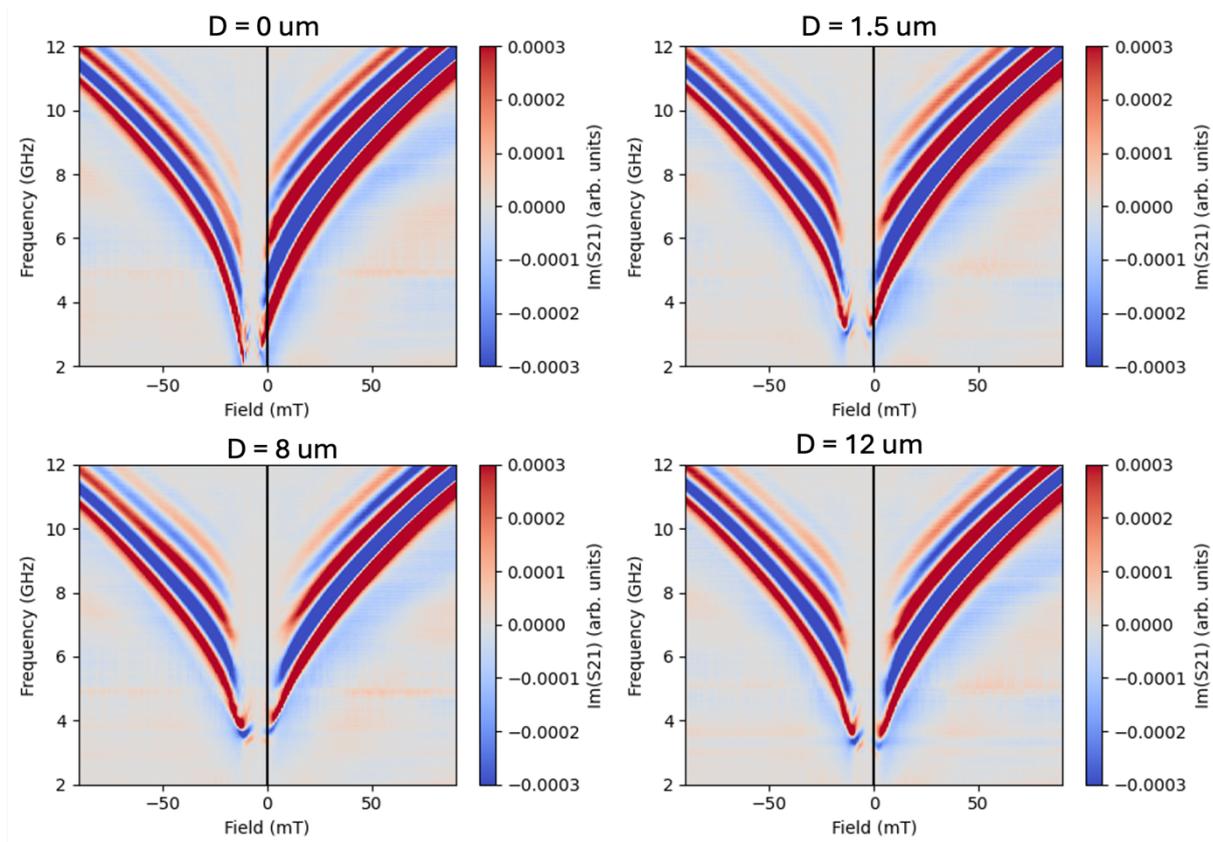

**Figure S5:** 2D color map of the oscillations related to the imaginary part of $S_{12}$ vs frequency for various applied fields Ha, for devices with varying distance D between magnets and MFCs.